# Start-up Shear of Concentrated Colloidal Hard Spheres: Stresses, Dynamics and Structure

N. Koumakis[1], M. Laurati[2], A.R. Jacob[1], K. J. Mutch[2], A. Abdellali[1], A. B. Schofield[3], S. U. Egelhaaf[2], J. F. Brady[4] and G. Petekidis[1]

[1]FORTH/IESL & Dept. of Mat. Sci. & Tech., Univ. of Crete, 71110, Heraklion, Greece

[2]Condensed Matter Physics Laboratory, Heinrich Heine University,

Universitätsstr.1, 40225 Düsseldorf, Germany.

[3]SUPA, School of Physics and Astronomy, The University of Edinburgh, Edinburgh, EH9 3JZ, UK

[4]Div. of Chem. and Chem. Eng., California Institute of Technology, Pasadena, CA 91125, USA

**Abstract**

The transient response of model hard sphere glasses is examined during the application of steady rate start-up shear using Brownian Dynamics (BD) simulations, experimental rheology and confocal microscopy. With increasing strain the glass initially exhibits an almost linear elastic stress increase, a stress peak at the yield point and then reaches a constant steady state. The stress overshoot has a non-monotonic dependence with Peclet number, Pe, and volume fraction, $\varphi$, determined by the available free volume and a competition between structural relaxation and shear advection. Examination of the structural properties under shear revealed an increasing anisotropic radial distribution function, $g(r)$, mostly in the velocity - gradient (xy) plane, which decreases after the stress peak with considerable anisotropy remaining in the steady-state. Low rates minimally distort the structure, while high rates show distortion with signatures of transient elongation. As a mechanism of storing energy, particles are trapped within a cage distorted more than Brownian relaxation allows, while at larger strains, stresses are relaxed as particles are forced out of the cage due to advection. Even in the steady state, intermediate super diffusion is observed at high rates and is a signature of the continuous breaking and reformation of cages under shear.

# I. INTRODUCTION

Colloidal hard spheres [1, 2] have been attracting much scientific attention as model systems for the experimental study of the glass transition [1, 3-5]. The phase diagram of this system is not affected by temperature but by particle volume fraction, φ. Intermediate volume fractions show crystallization, but at higher volume fractions (φ>0.59) they are kinetically trapped in a glassy metastable state where individual particles are spatially trapped by their neighbors and longtime diffusion is suppressed [2, 6, 7]. However, even in the glass regime some cooperative hopping processes may allow particles to escape, especially for samples near (but nominally above) the glass transition volume fraction leading to long-time restructuring and ageing [3, 8]. The linear rheological properties reflect the microscopic motions of the particles and dynamical arrest leads to a solid-like response which to a great extent can be described by MCT-based theories [9-14] although the low frequency relaxation mode detected in experiments at moderate volume fraction glasses is absent in ideal MCT [10].

With the application of shear above the yield stress or strain, irreversible changes and plastic deformation set in [15] leading to a nonlinear rheological response. The linear viscoelasticity and the steady rate response has been examined experimentally [9, 15-19], through theory and simulations [11, 20-23]. For concentrated glasses under steady shear, shear banding can occur mainly at low rates [24-26] with the shear banding instability for highly concentrated suspensions and glasses related to shear concentration coupling [25].

During a steady start-up shear experiment a transient response, featuring a linear increase of stress, in some cases a stress overshoot and finally, a steady state plateau is observed. The stress overshoot in a steady rate experiment contains information on the yield strain and stress of the system. Peaks such as these have been seen experimentally in a variety of soft matter systems such as polymers [27-29], worm-like micelles [30], nanocomposites [31, 32], colloidal gels and colloidal glasses [33-38]. Theoretical work exist based on Soft Glassy Rheology (SGR) [23] and Mode Coupling Theory (MCT) [11, 12, 39, 40] and contrasted with experimental data on hard sphere particles [41-43] and thermosensitive microgels [44], while overshoots have also been probed by simulations [30, 45, 46]. Although the precise explanations vary, the overshoot is a phenomenon stemming from an elastic energy storage mechanism and a dissipative energy release mechanism after the peak which fluidizes the system.

In order to link the mechanical response to changes in structure and dynamics we examine the microscopic motions of dense systems under shear. While initially arrested, the onset of flow for strains above the yield strain, allows for increase of microscopic particle motions due to out-of-cage diffusion. The use of confocal microscopy coupled with steady shear in colloidal glasses showed long time diffusive behavior of the non-affine motions, while additionally exhibiting a sub-linear power law dependence for the shear induced diffusion coefficient as a function of increasing shear rate [47]. However, the microscopic motions under shear have also been coupled to a modified Stokes-Einstein relation to replace thermal energy with shear energy [48].

In a range of strains similar to the stress overshoot, the transient dynamics have shown super-diffusive motion [42, 46, 49, 50]. While the stress peak has been associated with the appearance of the transient super diffusive behavior, recent work has found a strong correlation of the peak to local structural properties [49].

Concerning heterogeneities during flow, localized irreversible shear transformation zones have been identified [51] and plastic events under steady shear have also been rigorously analyzed [52] as well as regions of enhanced mobility in the transient regime [53]. A shear-induced heterogeneity length scale depending on the applied shear rate has been described [54], while a connection between diffusivity and heterogeneities [55] has been found in a two dimensional system. Still, however, the detailed physical mechanisms responsible for yielding in colloidal systems and the way this is affected by inter-particle interactions and microscopic properties is not fully understood. A recent review [56] examines colloids under shear with respect to their rheology, structure and microscopic rearrangements.

In this paper, we extend our earlier work [49], by studying the transient response of hard sphere glasses in a start-up shear experiment using Brownian dynamics simulations, experimental rheology and confocal microscopy over an extended range of parameters and implementing new analyses of the results. The combination of these three techniques allows the rheological characterization of the system under shear, while directly providing information on the microstructure and dynamics. The paper focuses on the correlation of shear stress to microstructural changes, while additionally putting mean dynamic properties into perspective.

## II. EXPERIMENTAL SYSTEMS AND TECHNIQUES:

### A. Samples

Nearly hard-sphere particles used here consist of polymethylmethacrylate (PMMA) spheres sterically stabilized by a thin (≈10 nm) chemically grafted layer of poly-12-hydroxystearic acid (PHSA) chains [32, 57]. For confocal measurements only, particles were fluorescently labeled with nitrobenzoxadiazole (NBD). For the rheological measurements we used particles with radius of R=267nm and a polydispersity of 6% suspended in decalin (cis-trans mixture), while the particles used in ageing studies had a radius of R=183nm, a polydispersity σ≈12% and were suspended in an octadenece/bromonapthalene mixture to avoid evaporation and minimize van der Waals attractions. The particles used for confocal microscopy with $R_C$=770nm and polydispersity of 6% were suspended in a solvent mixture of cis-decalin and cyclo-Bromoheptane (with the addition of 4 mM tetrabutylammonumchloride) closely matching density and refractive index. The size characterization was carried out with static light scattering measurements. The bare Brownian time, referring to the dilute limit, $t_B^0 = R^2/D_0$, with $D_0 = K_B T / 6\pi\eta R$ the free diffusion coefficient, $K_B T$ the thermal energy and $\eta$ the solvent viscosity, for the small (R=183 nm), medium (R=267 nm) and large (R=770 nm) particles are $t_B^0 = 0.13\ s$, $t_B^0 = 0.3\ s$ and $t_B^0 = 4.7\ s$, respectively.

The volume fraction for the hard sphere samples used in rheology were determined in two ways. For the relatively monodisperse sample (R=267 nm), a sample in the coexistence regime was prepared and the value of the volume fraction extracted from the liquid to crystal ratio of the sample [1]. For the more polydisperse sample (R=183 nm), φ was determined from random close packing [58], (which was assumed to be 0.67), which, however, carries a larger error due to the uncertainty in the polydispersity. Additionally, the packing fraction for a specific sample batch fluctuates with each packing/centrifugation around an average volume fraction, having a standard deviation of approximately σ_φ=±0.006, as determined in this work by using sequential centrifugations on the same batch. The volume fractions in confocal microscopy were measured through imaging [42, 59]. After calibrating a sample batch from coexistence, rcp or through imaging, the rest of the sample concentrations were determined by successive dilutions of the same sample batch.

### B. Rheology

The rheometer used for mechanical measurements was a stress controlled Anton-Paar MCR-501 with a Peltier temperature stabilization at T=20±0.01 °C. The area around the sample was sealed with a homemade vapor saturation trap. The smallest acquisition time used for this rheometer in the transient tests was 0.1 s to minimize error. Hence for the step rate tests low strain information was lost when the inverse rate was comparable or lower than the acquisition time. Two cone and plate geometries of different diameter where used to allow accurate measurements of low stress samples (large cone) and avoid thickening when loading for high volume fraction samples (small cone). The large cone was 50 mm in diameter with an angle of 0.0175 rad, while the small cone was 25 mm in diameter with an angle of 0.035 rad.

Before each measurement, a rejuvenation protocol was followed consisting of a series of tests starting with a high shear rate (10 s$^{-1}$) for 50 s, a small waiting time (30 s), the same shear rate but in the opposite direction for another 50 s and, usually, a waiting time of 200s, unless otherwise stated.

**C. Confocal Microscopy**

Confocal microscopy measurements were conducted in conjunction with a home-made parallel-plate shear-cell similar to those used previously [46, 60, 61]. In order to prevent wall slip the cover slips were coated with a layer of polydisperse colloidal particles [47, 62]. The rejuvenation protocol typically consisted of applying 10 large amplitude oscillations ($\gamma$>100%) at a frequency below 0.1 Hz and a waiting time of 600 s before each experiment.

Confocal Microscopy experiments were performed using a fast-scanning VT-Eye confocal microscope (Visitech International) mounted on a Nikon TE2000-U inverted microscope. A Nikon Plan Apo VC 100×oil immersion objective was used for all experiments. Two-dimensional images of the samples were recorded at a depth of 30 μm inside the sample in order to avoid boundary effects and to retain good quality images. The image size was chosen to be 512×512 pixels, corresponding to an area of 57×57 μm$^2$. In addition to slices, also volumes were imaged that consisted of 68 slices with a separation of 0.15 μm.

In a typical experiment, a series of images was recorded at fixed frame rate (typically much shorter than the inverse of the applied shear rate) starting simultaneously with the application of shear and ending when the maximum strain was reached. Particle coordinates and particle trajectories were extracted from images using standard routines [63]. Since particles do not move far between consecutive frames, it was not necessary to remove affine motions [64]. We typically averaged the results from 10 repetitions with about 1200 particles per measurement.

**D. Computer Simulations**

Brownian Dynamics (BD) simulations were conducted on hard sphere systems. The size of colloidal particles is such that there is a clear separation of time and length scales between the particles and the fluid molecules, therefore the fluid can be treated as a continuum, but the particles are still small enough to be affected by collisions with the fluid molecules and are thus still Brownian. Here we choose to use Brownian Dynamics simulations where Hydrodynamic Interactions (HI) between particles are ignored [65]. This allows simulations of larger and more concentrated systems, in comparison to the more accurate but computationally demanding Stokesian Dynamics (SD) simulation [65] where the full HI is computed.

For $N$ rigid particles of radius $R$ and density $\rho$ in a medium of viscosity $\eta$ moving with velocity $U$, we examine states where the Reynolds number (the dimensionless ratio of inertial forces $\rho U^2/R$ to viscous forces $\eta U/R^2$) is $Re \ll 1$. The motion of the particles is described by the N-body Langevin equation: $\mathbf{m}(d\mathbf{U}/dt) = \mathbf{F}^H + \mathbf{F}^B + \mathbf{F}^P$, where $\mathbf{m}$ is the generalized mass/moment tensor, $\mathbf{U}$ is the particle translational/rotational velocity vector, $\mathbf{F}^H$ is the hydrodynamic force vector, $\mathbf{F}^B$ is the stochastic force vector that gives rise to Brownian motion, and $\mathbf{F}^P$ is the deterministic non-hydrodynamic force vector. Since inertia is not important in colloidal dispersions ($Re \ll 1$) the equation reduces to $0 = \mathbf{F}^H + \mathbf{F}^B + \mathbf{F}^P$. For BD where HI between particles are neglected, the hydrodynamic force reduces to Stokes drag $\mathbf{F}^H = -6\pi\eta R \mathbf{U}$. The non-hydrodynamic force vector for a simple hard sphere system becomes the hard sphere interaction occurring at contact $\mathbf{F}^P = \mathbf{F}^{HS}$.

The hard-sphere interactions are calculated through the "potential-free" algorithm of [65, 66] in which the overlap between pairs of particles is corrected by moving the particles with equal force along the line of centres, back to contact. In order to calculate the stress, the algorithm directly calculates the pairwise inter-particle forces that would have resulted in the hard sphere displacements during the course of a time step [67]. Therefore we have $\mathbf{F}^P = 6\pi\eta R \left(\Delta x^{HS}/\Delta t\right)$, the average Stokes drag on the particle during the course of the hard-sphere displacement. Once the inter-particle forces from each collision are known, they can be used to calculate the stress tensor, $\langle \mathbf{\Sigma} \rangle = -n \langle \mathbf{xF}^P \rangle$ [67], where $n$ is the number density of particles and the angle brackets denote an average over all particles in the simulation cell. We should note that at rest and for relatively low Pe, Brownian Dynamics simulations should be able to qualitatively capture experimental stresses and particle motions, although hydrodynamics are neglected [65, 68]. However, stresses cannot be quantitatively compared.

By using BD simulations we attempt to elucidate the microscopic changes, both structural and dynamic, that occur during start-up shear. Various volume fractions, $0.54 \leq \varphi \leq 0.62$, with N=5405 particles were examined using multiple runs (typically 8 runs) by averaging with initial configurations, both in the glassy regime and below. We use a time step of $10^{-4}$ $t_B$ for Pe<1 (and rest), which scales to $10^{-4}$ per unit strain for Pe>1 [65]. Initial configurations were constructed by quenching from a dilute liquid state by appropriately increasing the particle size. The system was then properly equilibrated at rest allowing for a steady state to be reached by following the osmotic pressure and particle mean square displacements until both remained stable (typically after about 50 $t_B$).

This work mainly focuses and examines details on volume fractions, near the glass transition ($\varphi$=0.58) and well within the glass ($\varphi$=0.62) at rates of Pe=0.1, 1 and 10. In order to have greater clarity of the structural information, polydispersity was added in the simulations by a discrete Gaussian distribution of radii with a root mean squared deviation of 10%. The effect of the added polydispersity on stresses and displacements was found to be minimal. Moreover, due to the finite polydispersity crystallization under shear (or at rest) was not detected in experiments or in BD simulations.

**E. Analysis of Confocal Microscopy and Simulation Data**

Using the positions and trajectories of the particles, taken from simulations or microscopy, we calculate various structural and dynamical properties. The pair distribution function (PDF) describes the distribution of distances between pairs of particles contained within a given volume. We quantify structural anisotropy under shear by taking the projection of the radial-distribution function, g(r), in different planes relative to shear flow.

The Mean Squared Displacement (MSD) is a statistical measure of the distance a particle has moved in a specific time. If $x_i$ is the position of a particle i in x direction and N is the total number of particles then the MSD is calculated as:

$$\langle \Delta x^2(\tau) \rangle_{N,t} = \left\langle \frac{1}{N} \sum_{i=1}^{N} [x_i(t+\tau) - x_i(t)]^2 \right\rangle_t \quad (1)$$

where $x_i$ may be substituted with $y_i$ or $z_i$ for the other axes. In the case of the BD simulations, $x_i$ is the position of the particle calculated after the subtraction of the affine motion due to shear. It is important to note that by subtracting the affine motion at every time step, the effect of Taylor dispersion [69] is eliminated. Note that in the case of confocal microscopy the MSD is calculated in vorticity direction and thus no correction for the affine motion is required.

If the system is not in the steady state, the conventional time averaged MSD cannot provide information on the transient dynamics and the related microstructural changes. Thus we calculate two-time particle displacements using the equation:

$$\langle \Delta x^2(\tau, t_w) \rangle_N = \frac{1}{N} \sum_{i=1}^{N} \left[ (x_i(t_w + \tau) - x_i(t_w)) \right]^2 \quad (2)$$

which gives the average displacement between two times, with the time $t_w$ (>0) elapsed from the beginning of shear to the beginning of the measurement and t is the time elapsed since the beginning of shear with $\tau$=t-$t_w$ hence the time delay of the displacement. In this case, there is no averaging over time but only over the number of particles N and multiple measurements. Instead of $t_w$ and t, $\gamma_w$ and $\gamma$ may equivalently be used, denoting the acquired strain instead of elapsed time. Note that here the waiting time, $t_w$, has a different meaning from what is usually described in ageing experiments, namely the elapsed time after rejuvenation.

Moreover, we calculate the non-Gaussian parameter in the z direction:

$$a_{2z}(\tau) = \left(\langle \Delta z^4(\tau) \rangle_{N,t} \big/ 3\langle \Delta z^2(\tau) \rangle_{N,t}^2 \right) - 1 \quad (3)$$

which is the lowest order deviation of the Van Hove function from Gaussian behavior (averaged over particles and time). Non-zero values of the $a_2$ parameter correspond to non-Gaussian behavior and have been associated with dynamic heterogeneities [5], although, as discussed below, non-Gaussian behavior may not always be related with dynamic heterogeneities.

**F. Time-scale in experiment and simulations**

Both in experiments and simulations the dimensionless Peclet number, $Pe = \dot{\gamma} t_B$, was used to indicate the relative importance of shear convection relative to the diffusive Brownian motion. In the case of dense hard sphere suspensions the short time self-diffusion, $D_s^s(\phi)$ is reduced due to hydrodynamic interactions [70] compared to $D_0$. Since the actual Brownian time of the sample scales all other resulting times, when examining the effects of shear, the hydrodynamic effect on time scales should be taken into account. In the present work, we use the scaled Peclet number ($Pe_{sc}$) calculated with the volume fraction dependent short-time self-diffusion coefficient, $D_s^s(\phi)$.

The BD simulations do not incorporate hydrodynamic interactions, therefore in order to compare the rheological experiments and the BD simulations at rest and under shear, a normalization of time scales is needed. While in experiments $D_s^s(\phi)$ decreases with volume fraction, in BD simulations the in-cage short-time diffusivity is constant in the absence of HIs. Therefore in order to properly compare experiments, for which $Pe_{sc}$ is the relevant parameter, with BD we used values for $D_s^s(\phi)/D_o$ from Stokesian Dynamic simulations where HI were fully incorporated [70]. In table 1, we show the $D_s^s(\phi)/D_o$ used to scale BD to experimental data where $Pe_o = (D_s^s(\phi)/D_o)Pe_{sc}$. Note however that these values are approximate as there might be some discrepancy in volume fraction and particle polydispersity between experiments and simulations. We provide the bare $Pe_0$ and scaled $Pe_{sc}$ for experiments, while the $Pe$ in BD simulations should be compared with $Pe_{sc}$ in experiments as discussed above.

| Φ | 0.542 | 0.560 | 0.587 | 0.595 | 0.600 | 0.614 |
|---|---|---|---|---|---|---|
| $D_s^s(\phi)/D_o$ | 0.148 | 0.131 | 0.105 | 0.097 | 0.092 | 0.077 |

Table 1: The volume fraction dependent self-diffusion $D_s^s(\phi)/D_o$ that links the experimental results to the simulation results with $Pe_o = (D_s^s(\phi)/D_o)Pe_{sc}$.

**III. RESULTS**

**A. Rheological Response**

**1. Linear Viscoelasticity**

Fig. 1(a) plots the elastic and viscous moduli, G' and G'', in the linear regime for different volume fractions normalized with particle size and thermal energy, against frequency, ω, or equivalently $Pe_0^\omega = \omega t_B^0$. All samples shown here exhibit a solid-like, elastic response with G'>G'' in the frequency window examined. G' exhibits a weak (stronger) increase with $Pe_0^\omega$, especially for the higher (lower) φ's, while G'' shows a characteristic minimum, which moves to smaller times as φ increases, in agreement with previous findings [13, 38]. For the smallest φ = 0.542, G' approaches G'' at the low frequencies indicating an approach to a terminal flow regime where liquid-like response (G''>G') at lower frequencies is expected. According to the volume fraction dependence and the occurring regimes of G' and the minimum of G'', presented recently [38], the current system shows rheological indications of a glass transition (solid like response of the viscoelastic spectra with no indication of terminal flow) around φ=0.59. As recently discussed [59], the absolute volume fraction may entail an error of about 1%, however the relative volume fractions should be quite accurate.

## 2. Steady state stresses

In Fig. 1(b) we show flow curves, measured from high to low shear rates, for the different φ's. Each shear rate was averaged over 10 s to minimize the overall time of the test; accordingly the smallest rates ($Pe_0$<0.01) may not show the steady state response as the accumulated strain for each point is comparable to the yield strain. Shear banding (or slip) is not expected to be present in these measurements as we probe shear rates and φ that are in a stable flow regime [25].

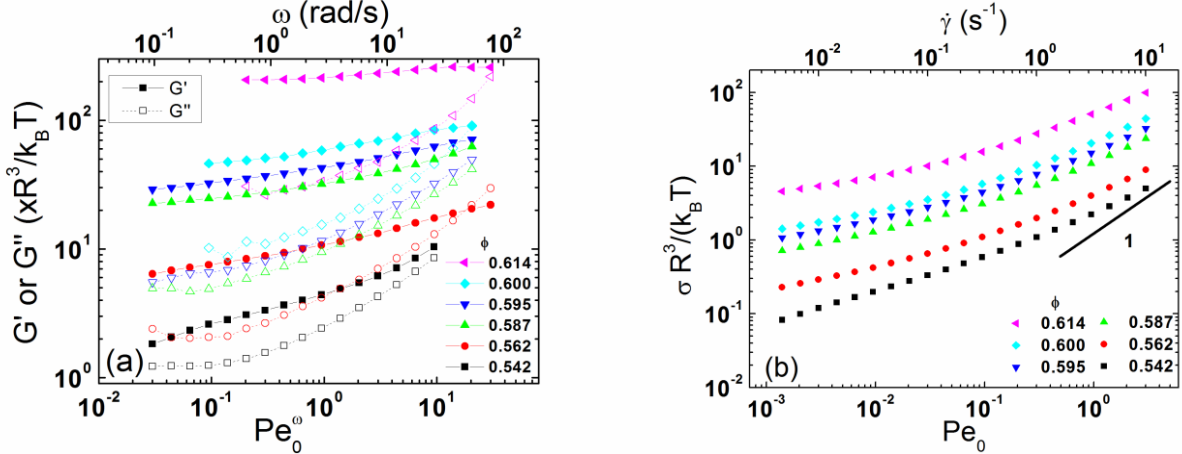

FIG. 1: (a) Dynamic frequency sweeps at different φ's as indicated for the R=267 nm PMMA particles in decalin. Solid symbols represent the elastic moduli, $G'$, and open symbols the viscous moduli, $G''$. (b) Corresponding experimental flow curves of various examined φ depicted in the legend. Line shows a power law slope of one. Horizontal axes are shown in the dimensionless $Pe_0^\omega = \omega t_B^0$ and $Pe_0$ (bottom) as well as the applied frequency, $\omega$, and shear rate, $\dot{\gamma}$ (top).

At high φ there is Bingham like behavior with a yield stress plateau emerging at low rates [16, 25]. According to simulations [71], for shear rates higher than in the experimental window, a slope of unity may be reached if shear thickening does not set in [72]. The lowest φ sample (0.542) shows some indication of the existence of a flow regime (power law slope approaching 1) both at high and low rates. The latter is related with the long time flow of the sample seen in linear oscillatory data where a G'/G'' crossover is expected for $Pe_0^\omega$<0.03 and terminal flow at even lower frequencies.

## 3. Transient stresses

Fig. 2 shows the measured stress versus strain in step-rate experiments for various φ at a fixed rate corresponding to $Pe_0$=0.0019. Since we examine various volume fractions, this corresponds to 0.013<$Pe_{sc}$<0.025. The experiments show an initial increase starting from a finite value, a peak in most cases and a steady state plateau for larger strains (γ>50%). It appears that the overshoot decreases with increasing φ. As mentioned above, the initial stress increase reflects an energy storage mechanism related to the entropic elasticity of the cage, while the peak is caused by a subsequent energy dissipation mechanism that leads to plastic flow and the plateau stress. Assuming a simple elastic response, the initial increase should show a linear increase with time (or strain), or equivalently a power slope of unity. For the data in Fig. 2 this only holds for small strains with the slope becoming smaller than one for larger strains. Furthermore, the stress curves have finite values of $\sigma$ for γ going to zero, as shown in the inset of Fig. 2 for one volume fraction (φ=0.62). This behavior indicates that the viscous component in the HS suspensions and glasses is significant for the corresponding time scales.

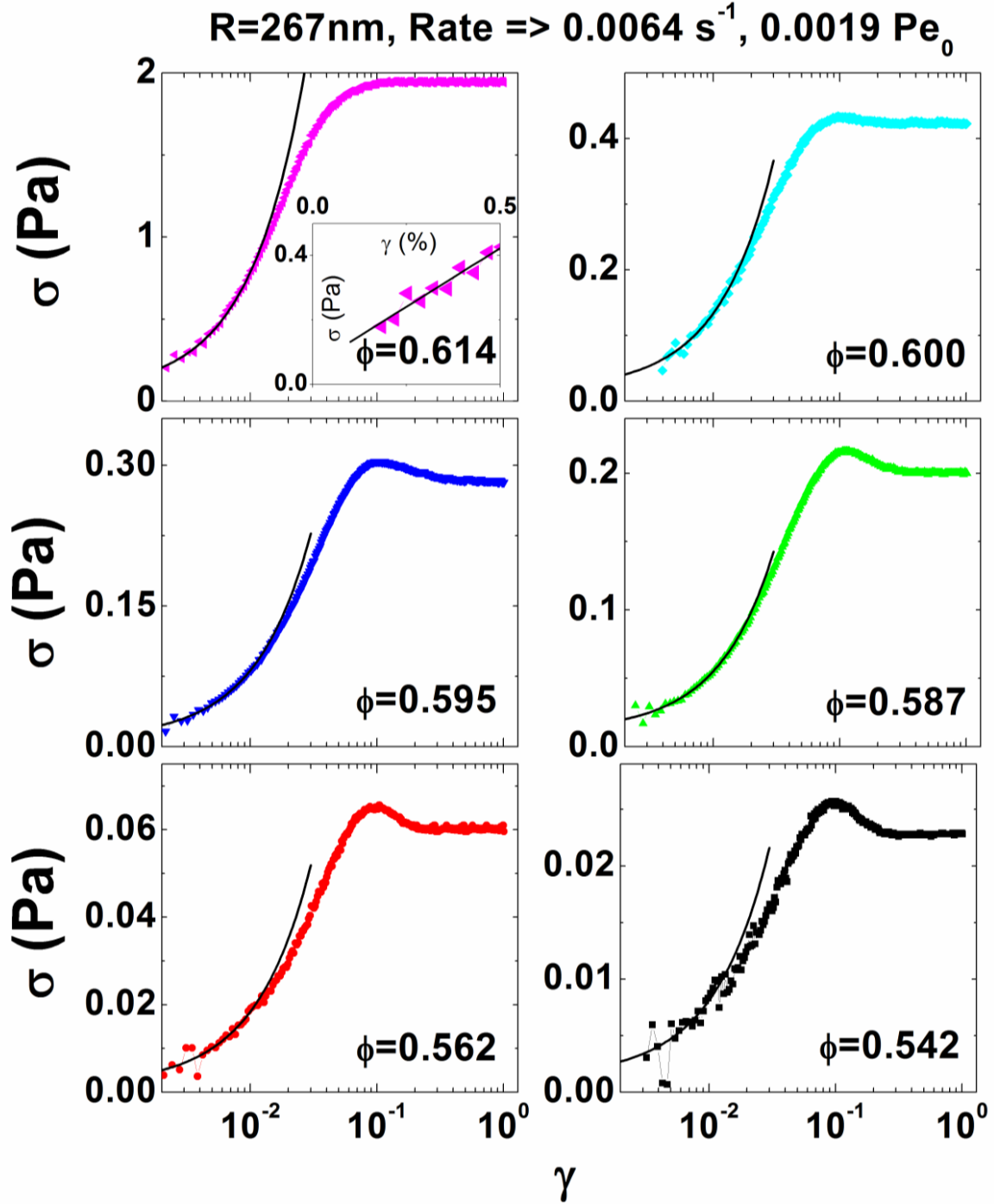

FIG 2: Stress, $\sigma$, versus strain, $\gamma$, in a lin-log plot from step rate start-up experiments of multiple φ for the R=267 nm PMMA particles in decalin, at a fixed rate resulting in $Pe_0=0.0019$ (or $\dot{\gamma}=0.0064 s^{-1}$) corresponding to $0.013<Pe_{sc}<0.025$. Lines are simple linear fits up to γ=1% as discussed in the text, extended to γ=3% to show the deviation from linearity. Inset of top-left shows the low strain data in lin-lin plot to indicate a nozero intercept. Measurements were taken at a waiting time of 200s ($t/t_B^0 \approx 667$)

The initial stress increase, up to about γ=1-2%, is given by linear response theory, although technically this is not straightforward to measure. In order to visualize the extent of linearity and the linear viscous contribution to the start-up experiments, we can utilize a simple equation for the viscoelasticity in the linear regime, $\sigma = \sigma_{elast} + \sigma_{visc} = G\gamma + \eta\dot{\gamma}$, under the assumption that during the examined experimental time window, the viscoelasticity (G, η) is constant. In a log-lin plot, the viscous contribution is apparent due to a non-zero stress at the start of the step rate experiments. Equivalently in a log-log plot, if $\dot{\gamma}$ remains constant then $\frac{d\log(\sigma)}{d\log(\gamma)} = \frac{G\gamma}{\sigma}$, which means that the power law slope in each point of the step rate, together with the total stress $\sigma$, can be used to decompose the total stress $\sigma$ into elastic, $\sigma_{elas}$, and viscous, $\sigma_{visc}$, components. Thus any step-rate experiment with an initial power law slope of less than one infers a finite viscous contribution to stresses. FIG 2 shows such fits to the data up to 1%, with extension of the curves up to 3%, thus additionally showing the deviation from linearity. The strongest deviation from linearity is seen for the highest φ, indicative of a smaller linear regime. Fit parameters G and η, return values close to G' and G''/ω (with the elasticity within 15%) as measured in the frequency sweeps at time scales corresponding to an accumulated strain of 1% for each rate.

Fig. 3 shows step rates tests at φ=0.587 for different shear rates. The values of the peak and the plateau stresses, as well as the strain at which the stress overshoot takes place are defined as $\sigma_{pk}$, $\sigma_{pl}$ and $\gamma_{pk}$ respectively. An increase of the shear-rate causes a rise of the stress, both at the peak, $\sigma_{pk}$, and at the plateau, $\sigma_{pl}$, while $\gamma_{pk}$ moves to higher values. The detailed shear rate and volume fraction dependence of the strength and position of the stress overshoot and the mechanisms relating it with the microscopic structure and particle dynamics will be discussed below.

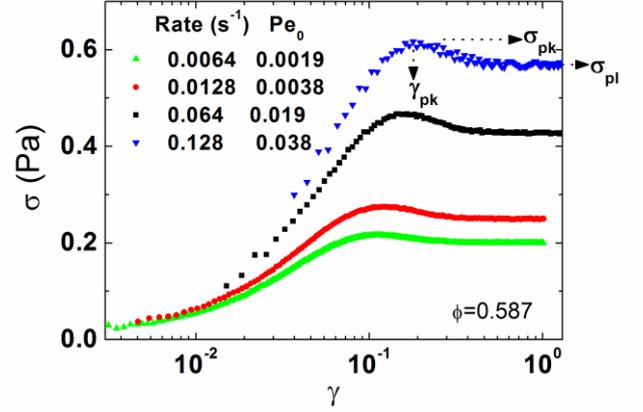

FIG. 3: Stress, $\sigma$, vs strain, $\gamma$, in a lin-log plot from step rate start-up experiments of φ=0.587 for the R=267 nm PMMA particles in decalin, at different rates as indicated, ranging between $0.0019 < Pe_0 < 0.038$ or equivalently $0.018 < Pe_{sc} < 0.366$. The values of $\sigma_{pk}$, $\sigma_{pl}$ and $\gamma_{pk}$ are also defined in the figure respectively as the peak and plateau stresses, as well as the peak strain. Measurements were taken at a waiting time of 200s ($t/t_B^0 \approx 667$)

The characteristic strain at the overshoot and their corresponding strength from experiments with constant ageing time of 200 s ($t/t_B^0 \approx 667$) are shown in Fig. 4. In Fig 4a, we see an increase of $\gamma_{pk}$ with shear rate as found previously [49] due to a stronger elongation of the cage before yielding, while $\gamma_{pk}$ is almost φ independent. Fig. 4(b) depicts the stress overshoot magnitude, $\sigma_{pl}/\sigma_{pk} - 1$, which exhibits an almost linear decrease with increasing φ at least for high volume fractions approaching random close packing. Note that at the highest volume fraction we do not detect a stress overshoot at some of the rates measured. The continuous decrease of the magnitude of the stress overshoot may be attributed to the diminishing free volume as the volume fraction is increased towards random close packing [49], which prohibits significant structural distortion and therefore stress storage and relaxation [49].

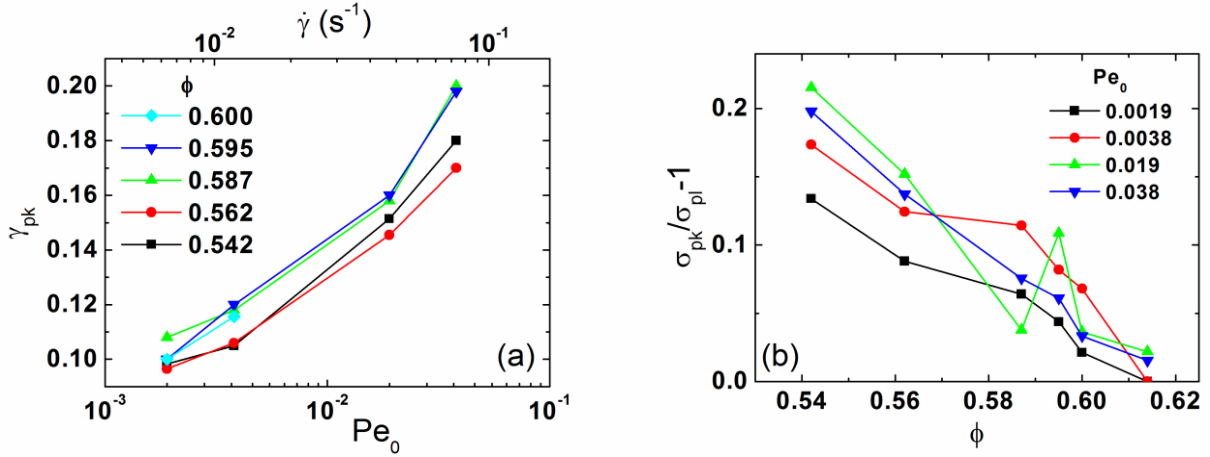

FIG. 4: Data from step rate experiments for the R=267 nm PMMA particles in decalin at an ageing time of 200s ($t/t_B^0 \approx 667$): (a) strain values of the peaks $\sigma_{pk}$ as seen in the Figs. 2 and 3 plotted against $Pe_0$ for different φ. (b) The normalized magnitude of the stress peak, $\sigma_{pk}/\sigma_{pl}-1$, plotted against φ for various rates indicating the loss of the peak at high φ.

## 4. Waiting time dependence (ageing)

Before we proceed further we briefly discuss the waiting time dependence (ageing) of the stress overshoot. The data presented in Figs. 2-4 as well as those in the main body of this paper were taken following a strict experimental protocol with a constant waiting of $t_{age}$=200 s (corresponding to $t/t_B^0 \approx 667$) after shear rejuvenation. This has proven sufficient to ensure relaxation to a reproducible steady state after rejuvenation, either using a protocol of zero shear rate or zero stress for the relaxation process, leading to invariant transient step rate data. It should be noted however that for longer waiting times, $t_{age}$, the strength, but not the position, of the overshoot is changing. Therefore, we have performed a detailed study of the effects of ageing on the stress overshoot using similar PMMA particles (with R=183 nm and σ≈12%) in an octadecene/bromonapthalene mixture to avoid evaporation and suppress van der Waals attractions. Although the full details of this study are beyond the scope of the present paper, here we show in Fig. 5 the resulting Pe dependence for different ageing times. As shown in Fig. 5 (a),(b) the stress overshoot becomes stronger with waiting time, mainly at low Pe, until it reaches a steady state at long times, often larger than 5000 s (corresponding to $t/t_B^0 \approx 3.89 \times 10^4$). At high Pe however (Fig. 5(c)) the stress response is almost independent of the waiting time. The increase of the stress overshoot with waiting time, at low Pe, is qualitatively similar to previous findings in Lennard-Jones glasses.

Fig. 5 reveals the two Pe regimes. At low Pe, as the shear rate is increased, the magnitude of the stress overshoot becomes stronger with waiting time. At long waiting times (here we reached beyond $t/t_B^0 \approx 10^4$) when the steady state is approached, the increase of the strength of the overshoot with shear rate is weaker. At high Pe however, Fig. 5 indicates a clear drop of the magnitude of the stress overshoot with the values affected much less by waiting time than at low Pe, if at all. Hence the magnitude of the stress overshoot exhibits a maximum at some characteristic Pe. This maximum in the response is more pronounced at shorter waiting times, while at long ones the increase at low rates tends to level-off. The characteristic Pe beyond which the stress overshoot starts to become weaker is slightly φ dependent. The same non-monotonic trend of the height of the overshoot as a function of Pe is seen for binary mixtures [73]

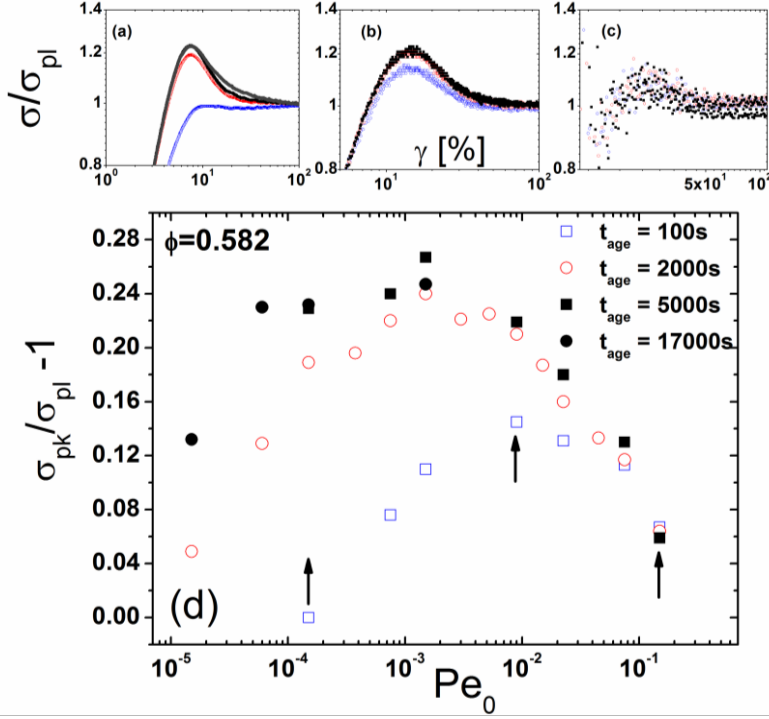

FIG. 5: Waiting time dependence of the stress overshoot for HS PMMA particles with R=183 nm at φ=0.575 in octadecene/bromonapthalene for different shear rates (Pe$_0$). (a, b, c): Normalized stress $\sigma/\sigma_{pl}$ versus strain $\gamma$ for start-up tests at three different shear rates ($\dot{\gamma}$ )$^{-4}$, $9\times10^{-3}$ and $0.15$), indicated by arrows in (d), and different waiting times as indicated by the different colors. (d): Strength of the stress overshoot $\sigma_{pk}/\sigma_{pl}-1$ as a function of Pe$_0$ for three different age times, $t_{age}$ =100s (blue open square), 2000s (red open circle) and the limit of long time with $t_{age}$=5000s (black solid square) or $t_{age}$ =17000s (black solid circle) (corresponding to $t/t_B^0 \approx 3.89\times10^4$ or $13.2\times10^4$ respectively).

In Fig. 6 the magnitude of the stress overshoot is shown as a function of volume fraction for different Pe$_0$. For an intermediate Pe$_0$ (=0.009) data are shown both for short and long waiting times for comparison. Ageing causes a change in the response at low rates. At short times ($t_{age} \approx$100s, corresponding to $t/t_B^0 \approx 778$) the stress overshoot drops continuously with volume fraction both below and above the glass transition volume fraction. At longer times approaching steady state a non-monotonic behavior is detected with a strengthening of the overshoot at low volume fractions and a weakening at higher ones. While this behavior is typically observed at low rates, at high ones (here Pe$_0$> 0.1) the magnitude of the overshoot decreases monotonically as φ is increased at all regimes both in the liquid and glassy state as found earlier and attributed to the decrease of free volume as the volume fraction is increased towards close packing [49].

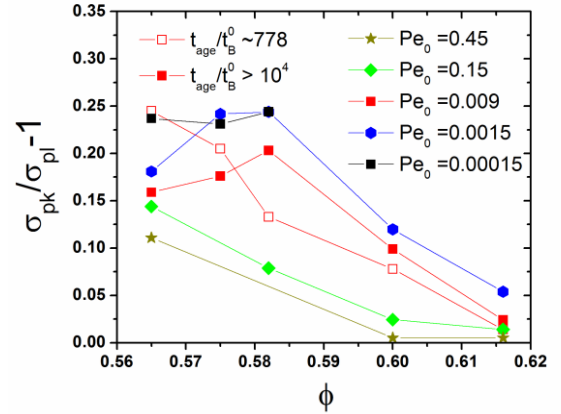

FIG. 6: Magnitude of the stress overshoot $\sigma_{pk}/\sigma_{pl}-1$ as a function of volume fraction φ (for the PMMA particles with R=183 nm in octadecene/bromonapthalene) at different shear rates quantified by the Peclet number Pe$_0$ (as indicated) in the limit of long ageing time (here for $t/t_B^0 \geq 10^4$). In addition for an intermediate Pe$_0$

(=0.009) data sets are shown for two age times as indicated.

## 5. Brownian Dynamics simulations

To complement the experimental rheological findings, we performed extensive Brownian Dynamics simulations at different volume fractions and shear rates probing the stress, structure and particle dynamics during start-up of shear. We first present the stress response (Fig. 7) for a start-up shear with Pe=0.1 and 1 at various volume fractions below and above the glass transition. The simulations reproduce stress overshoots similar to the experimental both in terms of the shear rate and the volume fraction dependence. Both $\gamma_{pk}$ and $\sigma_{pk}$ increase with Pe while the magnitude of the peak decreases as $\varphi$ is increased at high Pe. In order to quantitatively compare experiments and simulations the former were plotted as a function of $Pe_{sc}$. It should be pointed out that although this is possible at a single $\varphi$, the volume fraction dependences at a single shear rate involves comparison at different values of $Pe_{sc}$ as $t_B$ is $\varphi$ dependent.

Fig. 8(a) shows the characteristic strain $\sigma_{pk}$ of the stress overshoot from BD simulations as a function of Pe together with the corresponding experimental data plotted as function of $Pe_{sc}$. The qualitative findings are similar in both cases showing an increase of $\gamma_{pk}$ with increasing Pe.

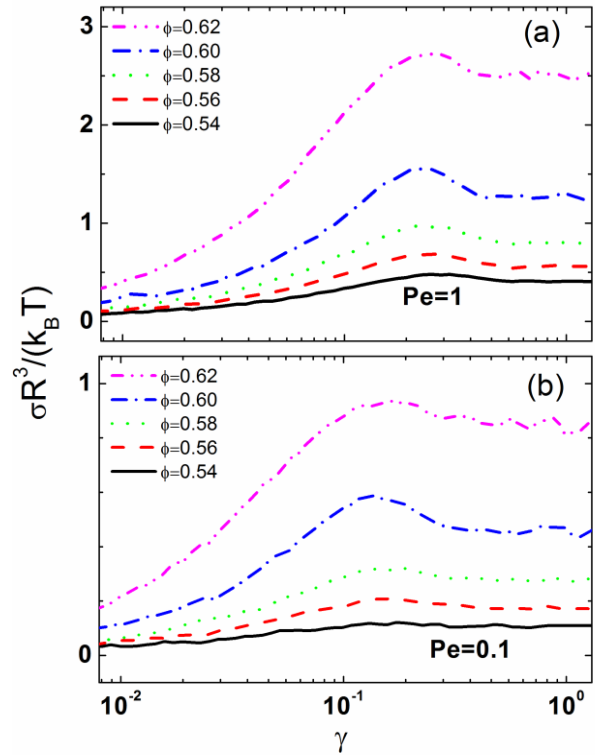

FIG. 7: Scaled stress, $\sigma R^3/K_B T$, versus strain, $\gamma$, in a lin-log plot from step rate start-up BD simulations for two different rates (a) Pe=1 and (b) Pe=0.1 at various $\varphi$ as indicated in the legend.

Fig. 8(b) depicts the $\varphi$ dependent magnitude of the stress overshoot from BD simulations and experiments. The trends observed in the latter at long waiting times are in accordance with findings in BD simulations. They reveal a complex volume fraction dependence that further depends on the Pe regime studied. At low Pe's the most notable feature of the $\varphi$ dependence is the existence of a maximum in the strength of the stress overshoot at a characteristic $\varphi$. The strength of the overshoot first increases with volume fraction with a slope that depends on Pe, being higher at low Pe (= 0.1 in BD) and weaker at intermediate (= 1 in BD). Above a critical $\varphi$ the stress overshoot starts to drop again and tends to diminish as the volume fraction is increased further in the glassy state as seen and discussed above for shorter waiting times. The critical volume fraction where this maximum of the stress overshoot takes place appears around $\varphi\sim0.6$ (Fig. 8b), but should be Pe dependent. At high Pe (= 10 in BD) the magnitude of the overshoot decreases constantly with φ below and above the glass transition in agreement with previous findings [49] and the long time, steady state experiments at high $Pe_0$ presented in Fig. 6. This suggests that at high Pe, where Brownian motion is no longer important, the cage structure appears static (no competition between Brownian motion and shear) and is deformed and subsequently broken under shear leading to relaxation of the stored stress. Since the stress overshoot reduction with increasing φ, is caused by the physical compaction of the cage structure (or first neighbor shell), it is reasonable to expect that the stress overshoot would also weaken and disappear, even at high Pe values, upon further lowering of φ in the liquid regime, although this is not yet obvious in Fig 8b. This would mean that at high, but still finite Pe, the critical φ where the stress overshoot is maximum is determined by the distance between first neighbors (size of the cage) and Pe which dictates if a particle can reach its cage limits due to shear before Brownian relaxation takes over. Some quantitative differences between BD and experiments observed in fig. 8b in the characteristic volume fraction where the maximum of the stress overshoot occurs could be due to the lack of hydrodynamic interactions, which may cause the simulations to underestimate the steady state stresses at lower φ. Otherwise, an absolute shift of φ and Pe between the two data sets may also be the reason.

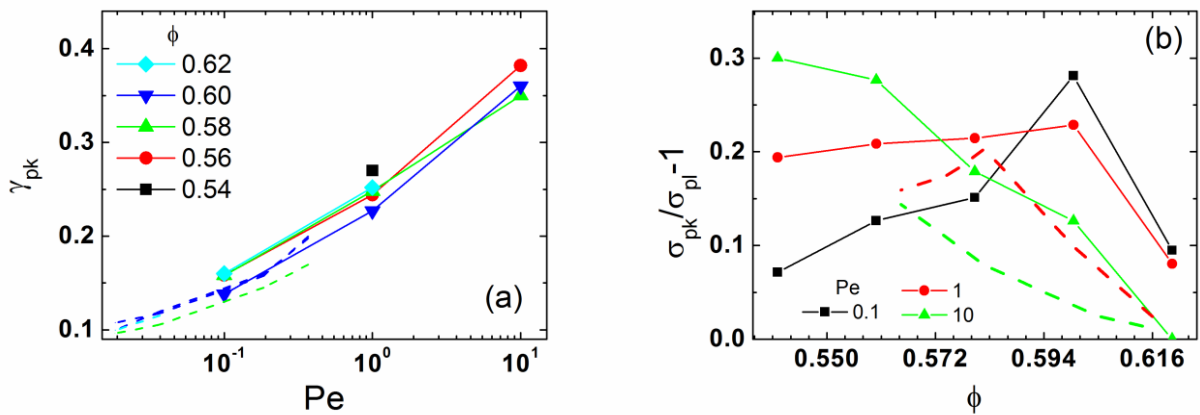

FIG. 8: (a) Strain values $\sigma_{pk}$ of the stress peaks as a function of Pe for various φ. Symbols connected by solid lines show the BD simulation results. Broken lines show experimental data for comparable values of $Pe_{sc}$ from Fig. 4 color coded to the nearest φ with BD simulations. (b) Magnitude of the stress overshoot, $\sigma_{pk}/\sigma_{pl}-1$, plotted against φ for a selection of rates indicating the loss of the peak at high Pe. Symbols connected by solid lines show the BD simulation results. Broken lines show experimental data from Fig. 6 at long ageing times for Pe=0.15 (broken green line) and for Pe=0.09 (broken red line).

### B. Structural properties and rheological response

#### 1. Transient Structure

In order to elucidate the origin of the stress peak in a hard sphere system, the transient and steady state structural information of the particles under shear [68, 72] is of great interest. The average radial particle distribution function, g(r), is unable to capture the anisotropy under shear, showing relative invariance from the state at rest. By examining what occurs at the three different planes xy (velocity-gradient), xz (velocity-vorticity) and zy (vorticity-gradient), a much clearer picture emerges. Since our interest lies in the local structure (cage), rather than the whole projection, g(r) data is gathered at a maximum distance of 0.7 radii from the plane. The resulting g(r) are 2D projections of the radial distribution function in the respective planes of slices with finite width (1.4 radii) and an area of 10x10 particle radii. The xy plane shows the most interesting features and its analysis gives an accurate description of the startup stresses. The values of the 2D g(r) are arbitrarily normalized to avoid clipping, although g(r) shown in the same figures have been normalized by the same factor. To achieve greater clarity for structural changes, from some g(r) the $g_{rest}(r)$ describing the state at rest was subtracted. g(r) are shown in false Blue-Green-Yellow colors, from low to high intensities, while g(r)-$g_{rest}(r)$ are in false Blue-Red colors, showing both negative and positive values.

Figs. 9(a) and (b) show 2D g(r) in the xy plane for φ=0.58 at rest and under shear (Pe=1) respectively. The g(r) correspond to a time average in the steady state. The xy plane anisotropy under shear presents itself as a higher intensity along the compression axis, indicated by a red line, and a reduction of intensity along the compression axis, indicated by a green line. The intensities are shown in more detail in Fig. 9(c), where the g(r) along the compression and extensional axes are compared to the isotropic condition at rest. Along the compression axis g(r) exhibits a first peak that is higher than that at rest, whereas it is weaker in the extension direction. At larger distances, the state at rest has the most well defined minima and maxima, followed by the compression and finally the extension axis. In contrast to the shifts of the positions of the first maxima, which are hardly discernible, the minima and higher order maxima do show a clear shift to smaller r/R in the compression axis and to larger r/R in the extension axis.

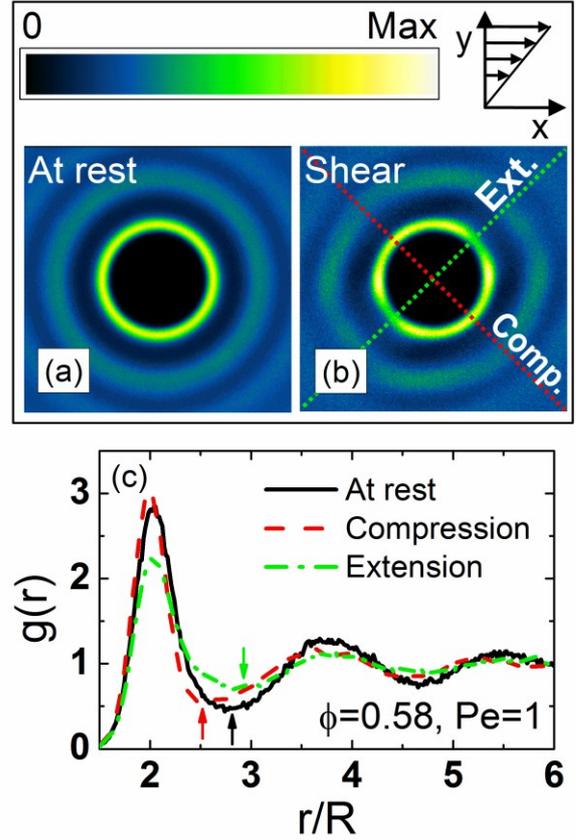

FIG. 9: Pair distribution function g(r) in the velocity-gradient (xy) plane as obtained by BD simulations for φ=0.58 (a) at rest and (b) at the steady state under shear at Pe=1 with the compression axis shown as a red line and the extension axis as a green line. The color scale along with the direction of the shear field is shown on the top. (c) Projection of g(r) for the same conditions along the compression axis, the extension axis and at rest. Arrows show the position of the first minimum of g(r) for the various directions.

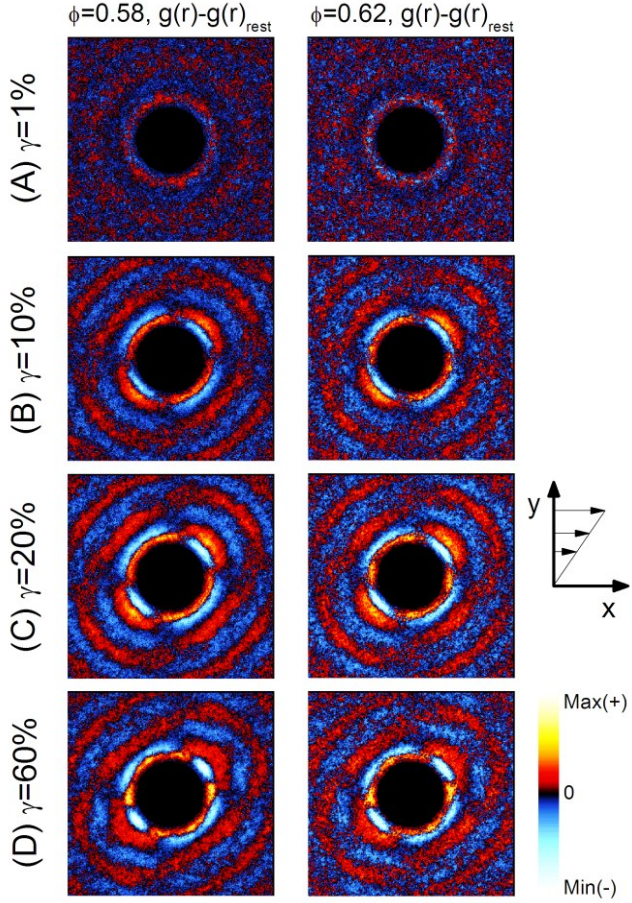

FIG. 10: Pair distribution function g(r) in the velocity-gradient (xy) plane after subtraction of the state at rest for φ=0.58 (left) and φ=0.62 (right) for Pe=1 at strains of 1%, 10%, 20% and 60%, from top to bottom, corresponding to the A, B, C and D positions in Fig. 11. The g(r) have been normalized with the same factor for all strains and φ.

The steady state g(r) shows that along the compression axis a large number of collisions occur, indicated by the increased height of the first peak and the positional shift of the subsequent minima and maxima. In the extension axis however, there are fewer collisions. Furthermore as the drop in the first maximum and the small positional shift to larger distances indicate there is a larger flow of particles that are in transit along the extension axis, "smearing" out g(r) in the extension axis.

The transient structural properties in the xy plane as well as the stress response for a startup shear simulation are shown in Figs. 10 and 11. Specifically, φ=0.58 and Pe=1 are chosen as a representative conditions due to the relatively large stress peak and clarity of the structural changes. Various points within the transient are examined, following the stress response in the linear regime at γ=1% (A), during the onset of nonlinear behavior at γ=10% (B), at the height of the peak, γ=20% (C), and at the steady state, γ=60% (D).

FIG. 10 shows 2D g(r) from φ=0.58 and 0.62 with the $g_{rest}(r)$ of the state at rest subtracted. As strain is increased, the anisotropy becomes more pronounced, starting from a small increase of the first maximum of g(r) in the compression axis at γ=1% (A). At γ=10% (B), the increase widens and anisotropy appears at the second peak, while the first maximum in the extension axis decreases and particle density increases at the first minimum. Increasing strain to 20% (C), the compression axis shows an even wider increase, with the extension axis showing a spread of particles reaching close to the secondary peak, while the intensity in the whole axis becomes more diffuse. In comparison to the peak stress at γ=20%, the steady state (γ=60%, D) reveals a g(r) with stronger intensity "lobes" in the x-axis of the first and second maxima, while the intensities of the first minima in the extension axis are reduced.

The observations from the g(r) of Fig. 10 are quantified in Fig. 11 which shows the first maximum values of g(r) as taken from the compression and extension axes, with the corresponding stress shown in the inset. The maximum of g(r) in the compression axis does not show much change with the application of shear. There is a clear increase due to an accumulation of particles in the compression axis but the resulting effect is small, since the state at rest is already at high particle density. However, the extension axis shows substantial differences from the state at rest, with the maximum of g(r) decreasing in amplitude with the application of shear and exhibiting a minimum at or slightly after the strain where the stress shows the peak. As shown in Figs. 9(c) and 10, the decrease of the g(r) maximum corresponds to the increase in probability of finding particles further away, towards the first minimum of g(r).

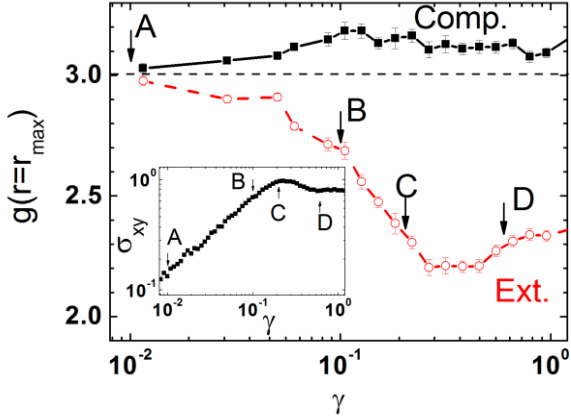

FIG.11: Maximum value of g(r) for the compression (■) and extension (o) axes (xy plane) as a function of strain for φ=0.58 and Pe=1 with the standard error of the average (8 runs). Inset depicts the transient stress $\sigma_{xy}$ versus strain $\gamma$. Positions A, B, C and D are marked at strains of 1%, 10%, 20% and 60% for both the main figure and the inset as indicated in Fig. 10.

Due to hard sphere interactions the maximum of g(r) directly reflects the stress response of the system. In general, the maximum of g(r) along the compression axis, $g_{comp}(r_{max})$ has a positive contribution to the $\sigma_{xy}$ stress component, while that in the extension axis, $g_{ext}(r_{max})$, a negative (Fig. 10). This results in a null stress response for an isotropic structure, and a non-zero stress for the anisotropy as caused by shear. Although it is not supposed to reproduce exactly the stress, the difference $g_{comp}(r_{max})-g_{ext}(r_{max})$ provides a qualitative measure of the additional stress built in the system due to shear induced structural anisotropy. In the case of these highly concentrated suspensions, the changes in $g_{comp}(r_{max})$, as shown above, are minimal and thus the contribution to stress is mostly due to changes on the extension axis.

By increasing φ and decreasing Pe towards Pe=0, the magnitude of the stress peak weakens, both in experiments (Fig. 4(b)) and simulations (Fig. 8(b)). Similar behavior is seen in the structural properties, $g_{comp}-g_{ext}$, as can be seen in Fig. 12, where the transient stresses for two φ (0.58, 0.62) and two Pe (0.1, 1) are plotted in tandem with the difference of the first maximum of g(r) in the compression and extension axes, $(g_{comp}(r_{max})-g_{ext}(r_{max}))$. The resulting maximum qualitatively shows similar behavior to the stress. Since the changes in the compression axis are minimal, the appearance and dependence of the stress peak are mostly correlated with structural changes in the extension axis.

The structural information during the transients and their correlation to the stresses provide strong insight on the origin of the stress peak in concentrated hard sphere systems. Interestingly, although counter-intuitive, both the increase of the maximum of g(r) on the compression axis and the decrease on the extension axis contribute positively to the shear stress. This is understood by recalling that the departure from rest of the relevant deviatoric stress is 
$$\sigma_{xy} = -n^2 k_B TR(2R)^2 g_0(2R)\iint n_x n_y f d\Omega \quad (4)$$

with n as mentioned above is the particle number density, $g(2R)=g_0(2R)(1+f)$, $g_0(2R)$ the value at rest, dΩ the solid angle and $n_x$, $n_y$ the components of the separation vector between two particles at contact in the x and y direction. Hence $\sigma_{xy}$ is positive both in the compression axis where f is positive and $n_x n_y$ negative and in the extension axis where the signs of f and $n_x n_y$ are reversed [49].

Consequently, as the stress is related with the number of particle collisions both those along the compression axis and those along the extension axis are additive. Since the first maximum of g(r) is proportional to the number of occurring collisions (higher particle density is tied to more frequent particle interactions) Fig. 12 indicates that the structural origin of the transient and steady stresses lies in the disturbance of the isotropic pressure balance, more specifically the reduction of collisions in the extension axis, rather than the increase of collisions in the compression axis.

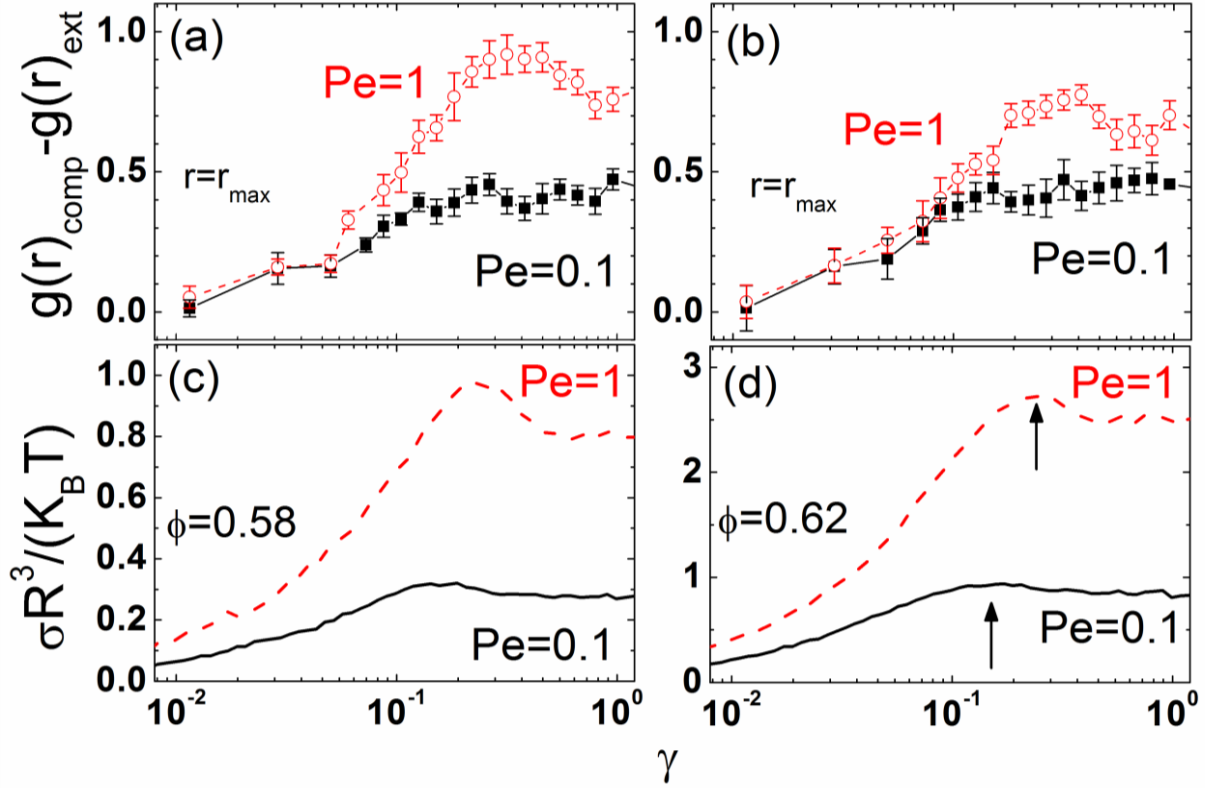

FIG. 12: Transient stress $\sigma R^3/K_B T$ versus strain $\gamma$ from simulations (bottom) shown in relation to the difference of the maxima in the pair distribution functions in compression and extension direction, $g_{comp}(r_{max})-g_{ext}(r_{max})$ as a function of strain (top) for $\varphi=0.58$, $0.62$ and Pe=0.1, 1. Error bars in (a) and (b) indicate the standard error for the average of 8 runs, while vertical arrows in (d) specify the position of the stress peaks.

As mentioned already in Fig. 11, we see that the peak stress takes place at or very close to the minimum of $g_{ext}(r_{max})$. More details on the peak can be acquired by examining the differences in the extension axis between the (C) and (D) points of Fig. 11. At 20% strain, (C), the extension axis shows a high concentration of particles in the first minimum of $g_{ext}(r)$, particles which are on their way to escape their cage, but are still trapped. Well beyond the stress overshoot, at 60% strain, (D), and onwards at steady state, this clustering of particles has decreased and is spread out, as particles are escaping the cage and flow takes place. Thus the minimum in $g_{ext}(r_{max})$ is a manifestation of the escape of particles initially trapped within the cage thereby reducing the stress.

Further understanding of the stress peak lies in the volume fraction and Pe dependence. As the shear rate is decreased towards Pe=0, both the strain and amplitude of the peak decrease (Figs. 3, 4, 5 and 12). At the lowest shear rates, Brownian motion dominates and particles which would be trapped at higher rates have time to escape through thermally activated diffusive motion, reducing the height and the strain of the peak. On the other hand, at higher rates, particle escape is activated by shear induced collisions, while the trapping of particles in an anisotropic structure at intermediate strains around the yield strain causes a stress overshoot.
17

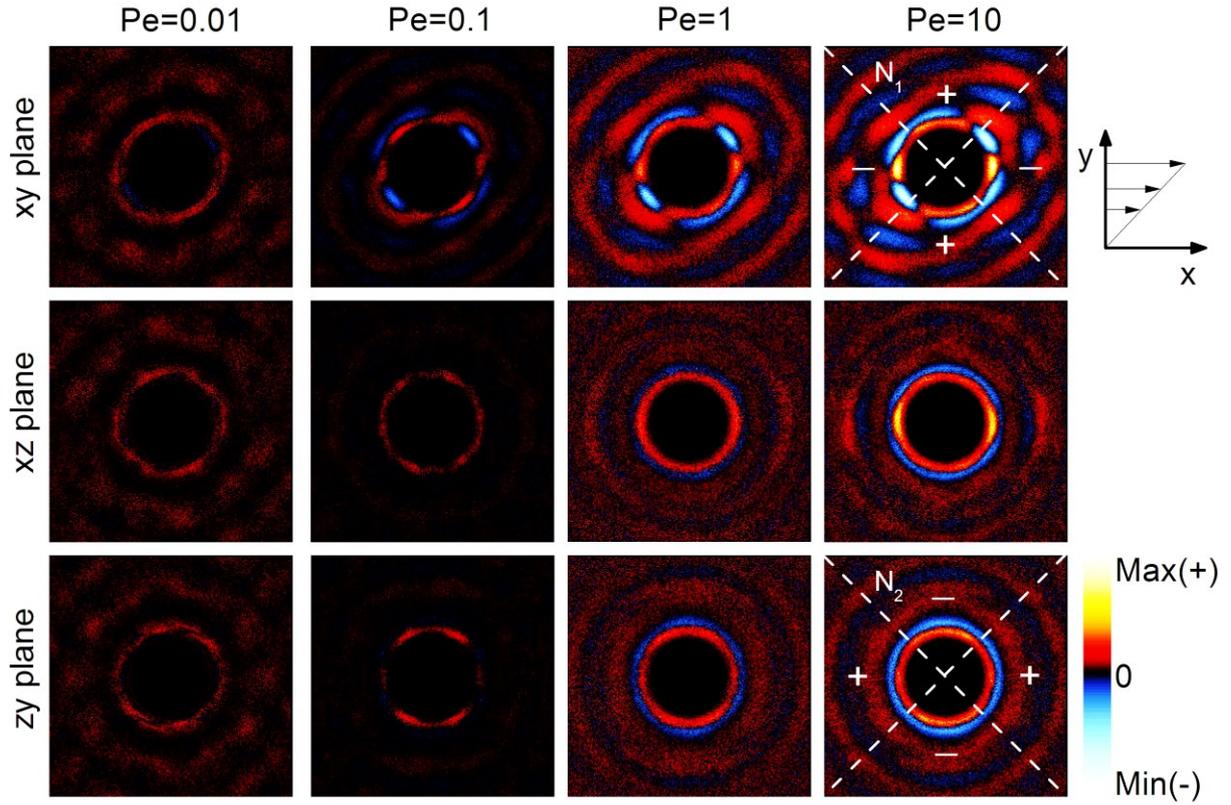

FIG. 13: Pair distribution function g(r) for φ=0.58 with Pe=0.01, 0.1, 1 and 10 (From left to right) in the velocity-gradient (xy), velocity-vorticity (xz) and vorticity-gradient (zy) planes (from top to bottom) averaged at the steady state and shown after subtraction of the g(r) at rest. The contributions of the various g(r) planes to normal stresses, $N_1$ and $N_2$, are selectively shown. All g(r) in the figure have been normalized by the same factor.

Therefore, with the application of high shear, the cages are able to store energy in the form of cage elongation or distortion, which is then released when the cage finally breaks. In the case of increasing φ, there is less space available for cage elongation, leading to weaker stress overshoots.

However we also have seen in Fig. 5 that when the shear rate exceeds a certain value, at high Pe, the strength of the overshoot starts decreasing again. This indicates that we have passed to the regime where Brownian motion is not significant any more (approaching the non-Brownian regime) where entropic elasticity becomes progressively less efficient in recoiling cage deformation after the yield point. Therefore in this regime, beyond a critical Pe that is volume fraction dependent the stress overshoot progressively weakens. This dependence is also seen in BD simulations at high volume fractions and maybe is related microscopically with collision induced cage escape of particles in a wider strain range around the yield point, a mechanism that is promoted at high rates and volume fractions. The latter is also supported by the fact that the critical Pe beyond which the stress overshoot decreases again shifts to lower values with increasing volume fractions as seen indirectly in Fig. 8(b) by comparing the curves at different Pe, and in agreement with findings in binary hard spheres [73]

The "lobes" occurring in the steady state under shear (γ>60%, Fig. 10) are due to the geometry of simple shear flow. Simple shear occurs through the addition of an extensional flow (in the extension axis), coupled with a rotational flow. The combination of the two leads to a higher probability for particle collisions along the shear direction, as well as in the compression axis.

## 2. Steady State Structure

Next we turn our attention to the structure deduced from BD simulation at steady state in all three different planes relative to shear. Fig. 13 shows the steady state g(r) (after subtracting that at rest) in the xy, xz and zy planes at Pe=0.01, 0.1, 1 and 10 for $\varphi$=0.58. Different rates show similar features when looking at the same plane, although with less intensity as Pe and thus shear induced cage deformation is decreased. Besides the features in the xy (velocity-gradient) plane discussed above we observed that g(r) in the xz (velocity-vorticity) shows an increase of the first and second maxima with a higher localized intensity along the x axis. The zy (vorticity-gradient) plane mirrors the features of the xy plane, with the localized intensity increasing in the y axis. The intensity variations found for Pe=0.01 are due to statistical errors in the structural differences.

## 3. Confocal Microscopy Stress Extraction

Confocal microscopy experiments with the nearly hard-sphere PMMA particles have produced similar transient structures as can be seen in Fig. 14 for $\varphi$=0.56 and $Pe_{sc}$=0.594. From the structural information we can estimate the stress tensor using the hydrodynamic pair particle approach [65, 68, 74], through equation

$$\langle \mathbf{\Sigma} \rangle \sim -n^2 k_B T \int_{r>2R} \frac{75}{2}\left(\frac{R}{r}\right)^6 \left(\frac{\mathbf{rr}}{r^2} - \frac{1}{3}\mathbf{I}\right) g(\mathbf{r}) d\mathbf{r} \quad (5)$$

where $\mathbf{r}$ is the particle separation vector and $\mathbf{I}$ the isotropic tensor. Since the equation is modeled on dilute hydrodynamic interactions, it can only qualitatively capture the stress response of the concentrated hard sphere suspension, in a similar fashion to the BD simulations.

Fig. 14 shows the stress calculated according to Eq. 5 from the structural information and particle coordinates measured in confocal microscopy experiments. Although measurements do not extend beyond a strain of 25% the stress response captures the main features measured in rheology showing an increase which then reaches a rough plateau. Since the volume fraction and shear rates are low in confocal microscopy, the stress peak is not strongly apparent, in agreement with rheological experiments and simulations (Fig. 8(b)). Nonetheless, confocal microscopy experiments verify the main structural findings from BD simulations and may be further used to quantitatively extract the stresses measured in macro-rheological measurements

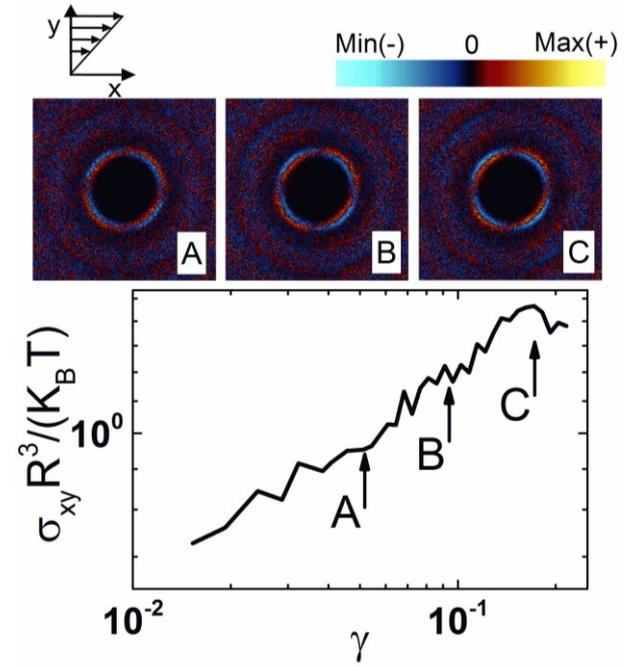

FIG. 14: Above: Pair distribution function g(r) in the velocity-gradient (xy) plane after subtraction of the state at rest for confocal measurements at $\varphi$=0.56 for $Pe_{sc}$=0.594 at strains of 5%, 9% and 17% corresponding to the A, B and C positions as shown in the below. Below: The stress calculated based on confocal micrographs as described in the text.

## 6. Transient normal stress measurements

As a final note with regard to the structure - rheology relation we present the behavior of the normal stresses as determined from BD simulations and confocal microscopy data during a start-up test. The first and second normal stress differences are given by $N_1=\sigma_{xx}-\sigma_{yy}$ and $N_2=\sigma_{yy}-\sigma_{zz}$, respectively where $\sigma_{xx}$, $\sigma_{yy}$ and $\sigma_{zz}$ are the diagonal components of the stress tensor with their average being the suspension osmotic pressure $\Pi=-(\sigma_{xx}+\sigma_{yy}+\sigma_{zz})/3$. In concentrated hard sphere suspensions normal stresses have non zero values which are both negative at high shear rates where Brownian motion is not important, while $N_1$ acquires positive values at low Pe where Brownian motion dominate [72].

Fig. 15(a) shows the transient $N_1$ and $N_2$ as a function of strain for Pe=0.1 and $\varphi$=0.56 from BD simulations, while Fig. 15(b) shows the same quantities extracted from confocal measurements for $Pe_{sc}$=0.594 and $\varphi$=0.56 (same data as Fig. 14). Although the general behavior observed is reminiscent of the shear stresses, $\sigma_{xy}$, we find that $N_1$ has positive values, while $N_2$ has negative values. The sign of the normal stress differences can be understood by examining the structure under shear. Fig. 13 schematically shows the relation of the g(r) intensity to $N_1$ and $N_2$ in the xy and zy planes (velocity-gradient and vorticity-gradient) as presented in [72]. In general, the increase of intensity at contact along a certain direction represents a larger pressure giving rise to an increase in the corresponding diagonal stress element in absolute values. Consequently, $N_1$ is positive and $N_2$ is negative as the g(r) at contact shows higher intensity along the y axis in the corresponding g(r) (xy and zy planes respectively). We should note that since these values are quite small the error is much higher than that involved in the determination of $\sigma_{xy}$. Nevertheless, for the simulations, a clear peak in $N_1$ and - $N_2$ is seen around the position of the stress overshoot. The nature of the normal stress transient peaks is of similar origin to that of the shear stress, i.e. the elongation of the cage before breakage.

**C. Microscopic Dynamics**

**1. Direction and Rates**

The physical description of yielding in a concentrated hard sphere system can be completed by examining microscopic particle motions under shear. Although this has been the focus of previous work [42], here we present a short overview of the relevant phenomenology and its physical interpretation in relation to the stresses and structure. Moreover we extend this study in presenting additional data providing information on the volume fraction dependence. In Fig. 16, we examine the transient and steady state displacements at high $\varphi$, along the compression and extension axes which was shown to be the most influential on the stress peak.

Fig. 16 shows particle mean square displacements under shear for BD in the compression (16a) and extension (16b) axes for $\varphi$=0.62 and three different Pe (0.01, 0.1 and 1) in comparison with the corresponding particle displacements at rest. For the lowest Pe=0.01, only a single run was examined due to long simulation times. In Fig. 16 (c) and (d) we also show the corresponding effective diffusivity, $D_{eff}=<\Delta x_i^2>/[2(t-t_w)]$. As mentioned earlier, the data under shear represents an average over all particles, N, during a time window of $\Box$=t-t$_w$, or equivalently an elapsed strain of $\Delta\gamma=\gamma-\gamma_w$. Two states are chosen: a) a transient state with $\gamma_w$=0, revealing the evolution of the particle displacements during the start-up test and b) a steady state with $\gamma_w$>0.5 where the displacements are monitored during well-developed shear flow. Note that corresponding stresses for these tests are shown in Fig. 12.

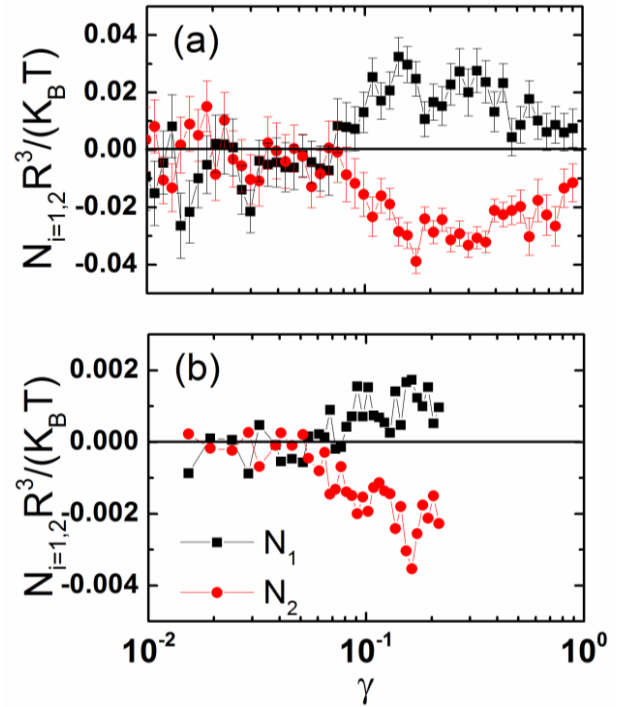

FIG. 15: The transient normal force components $N_1$ and $N_2$ from (a) simulations on Pe=0.1 for $\varphi$=0.56 and (b) confocal stresses at $\varphi$=0.56 for $Pe_{sc}$ = 0.594, extracted as discussed in the text. Error

bars in (a) show the standard error of the average from multiple runs (8 runs), while (b) shows a single experimental measurement.

For the transient state, $<\Delta x_i^2>$ generally shows an initial increase which follows the response at rest, with a subsequent transient super-diffusive behavior [42, 46, 49, 50] which eventually becomes diffusive (slope of unity) at longer t-$t_w$. At large $t_w$ (or $\gamma_w>0.5$), corresponding to the steady state of well-developed shear flow, $<\Delta x_i^2>$ matches the diffusive behavior for long t-$t_w$, whereas at short t-$t_w$, it shows lower values compared to the displacements at rest due to in-cage constriction of particles (Figs. 16(a) and (b)). The latter is better seen at high Pe, whereas at the lowest Pe, is essentially absent; moreover particle dynamics ($<\Delta x_i^2>$ or $D_{eff}$) for the transient and the steady steate are almost identical, thus not well discerned in fig 16. The transient super-diffusion and constriction effects can be seen more clearly in the Figs. 16(c) and (d) which show the effective diffusivity, $D_{eff}$. A decreasing effective diffusivity represents sub-diffusive motion, a constant value shows simple diffusion, while a positive slope reflects super-diffusive motion. As Pe is decreased, long-time diffusivity decreases, while both the constriction and super-diffusive regimes becoming less pronounced and eventually fading out at the lowest Pe=0.01. In Fig. 17(a), we show that these effects also diminish with decreasing volume fraction. By comparison of the dynamics on the compression (16a, 16c) and extension (16b, 16d) axes, we additionally observe that the effects of constriction at short times, the transient super-diffusion and steady state super-diffusion are all enhanced in the direction of the compression axis. Although not obvious by a first look a careful inspection of the long-time diffusivity depicted in Figs. 16 and 17 reveals that at steady state these are slightly, but clearly, different in various directions: the highest being in the extension axis, followed by that in the compression axis and finally the lowest measured in the vorticity axis.

The super-diffusive response, particularly for $t_w$=0, is possibly a reflection of the elastic, entropic recoil of a distorted cage. This however is manifested as a ballistic like motion at the very short times corresponding to length scales within the cage, as it takes place around $(<\Delta x_i^2>/R^2)^{1/2} \sim$ 0.1. This corroborates another important finding, namely that the position (strain) of the stress overshoot does not coincide with the minimum in the effective diffusivity representing the super-diffusive behavior. Instead, the relevant strains of super diffusivity are smaller than those where the stress overshoot is observed, ranging from about the beginning of nonlinearity up to the stress peak. Therefore, the transient super-diffusion can be interpreted as a ballistic-like, cooperative motion of the particles during yielding, as they are being pushed towards the limit of the cage due to shear, prior to yielding. Interestingly, the transient displacements along the extension axis deviate earlier from those at rest than the displacements along the compression axis. This suggests that yielding is initiated in the extension axis as also seen in the structural data of Figs. 10 and 11. Along these lines the super-diffusive motion may more specifically reflect a concerted particle transition from the compression to the extension axis as the structure is continuously deformed with more particles brought together along the compression axis and then escape out of the cage mainly on the extension axis. An analogous mechanism for shear induced particle rearrangements has been suggested to relate with structural distortions during similar start-up tests and a change of symmetry due to yielding [41].

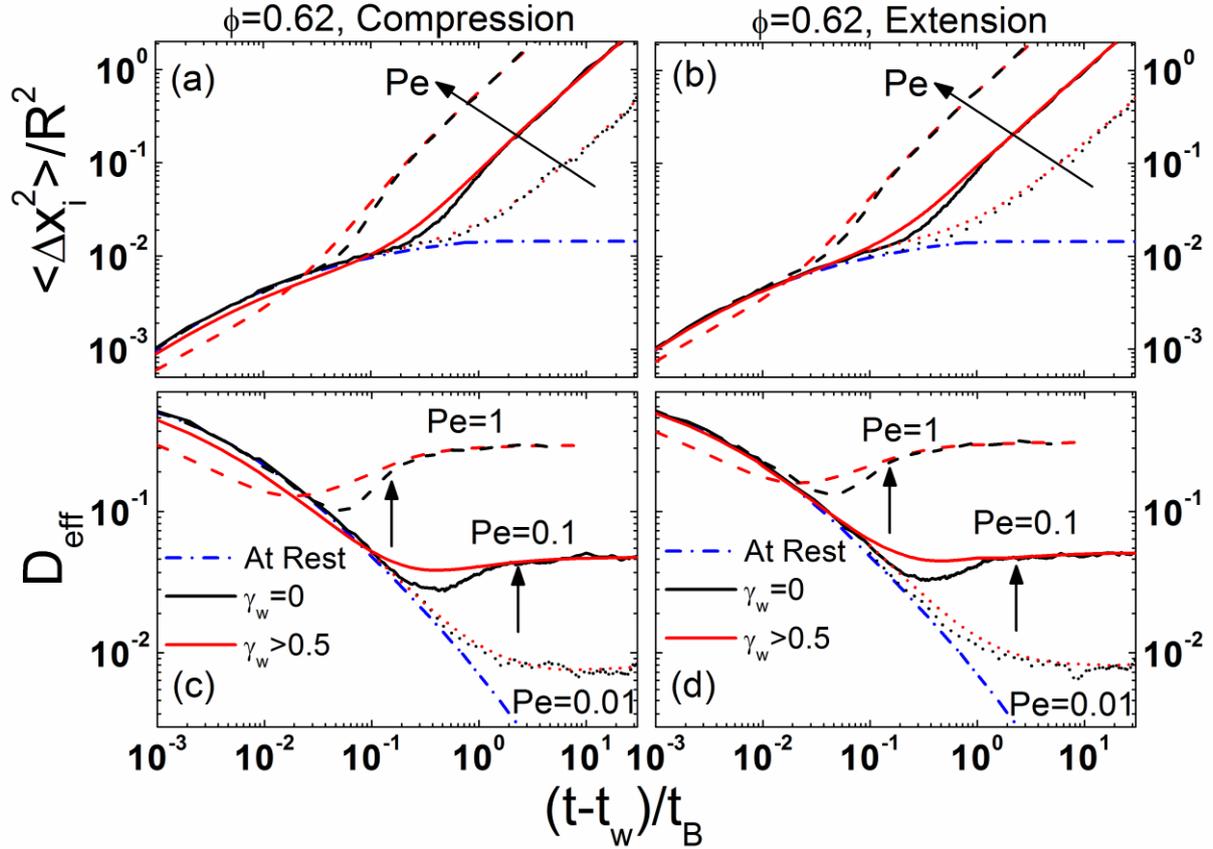

FIG. 16: Two time mean squared displacements for $\varphi=0.62$ in comparison to the state at rest, examined in the (a) compression and (b) extension directions from start-up shear simulations at Pe=0.01, 0.1 and 1, using waiting strains of $\gamma_w=0$ (transient) and $\gamma_w>0.5$ (steady state). The corresponding effective diffusivity, $D_{eff}=<\Delta x_i^2>/[2(t-t_w)]$, is shown for the (c) compression and (d) extension directions, with arrows showing the corresponding positions of the stress peaks.

The constriction of in-cage motion at short time scales during steady state, is found to be correlated to the distortion of the cage at high rates (Fig. 13). Such cage deformation leads to a smaller free volume per particle, and a subsequent decreased mobility. Thus, the effects of constriction and transient super-diffusion are more pronounced in glassy samples in the compression axis, as a consequence of the smaller cage dimension. A similar drop of short-time, or instantaneous, diffusivity with increasing shear rate has also been seen at lower volume fractions in the liquid regime by Stokesian Dynamics simulations [72], in confocal microscopy experiments on binary colloidal glasses [73] as well as in orthogonal superposition rheometry of colloidal glasses at high frequencies during shear [75].

A relevant interesting effect of *steady state* super-diffusion occurs at higher shear rates and $\varphi$ as can be clearly seen in the corresponding Figs. 16 (c) and (d) of the effective diffusivity for Pe=1. This phenomenon may be related to the enhancement of spatial constriction under these conditions, through a transition from highly constrained in-cage motions to an out-of-cage shear-induced diffusion. Although it takes place continuously beyond the stress overshoot the origin of the steady state super-diffusion should be similar to that of the transient super diffusion. In the latter case, as discussed above, particles transition from in-cage motion at rest to sheared out-of-cage diffusion through a mechanism that involves a concerted motion from the compression to the extension axis that gives rise to a transient effect which advances the stress overshoot (Fig. 16). In steady state super-diffusion, the cage constriction effect is constantly renewed especially at high shear rates where strongly deformed cages (or a large structural anisotropy) are established under flow. Therefore, in the transition from in-cage to out-of-cage motion particles exhibit a concerted motion from the compression to the extension axis. Such localized ballistic-like motion causing the steady state super-diffusion together with the out-of-cage (long-time) diffusivity, are the dynamic signatures of a sequence of structural events involving cage breaking and reformation in a colloidal glass under shear.

## 2. Volume fraction dependence

Of additional interest is the volume fraction dependence of shear-induced dynamics and their comparison to the dynamics at rest. Fig. 17 shows the $\varphi$ dependence of the transient and steady state effective diffusivity and the non-Gaussian steady state behaviour for an intermediate shear rate from BD simulations and confocal experiments in the vorticity direction. Simulations at Pe=0.1 (Fig. 17(a)) for a range of $\varphi$ from below (0.56) to above (0.62) the glass transition, show that the long-time diffusivities under shear decrease with $\varphi$. Confocal experimental data of Fig. 17(b) at $\varphi$=0.56 and 0.57 for $Pe_{sc} \approx 0.5$ indicate a consistent trend, although volume fraction values may be too close to discern a difference. The transient diffusivity minimum becomes more apparent and changes time scales for various $\varphi$, even though the strain of the peak remains approximately the same. Moreover, at short times a pronounced constriction effect is detected at higher $\varphi$ in BD simulations.

The change of the minimum of the effective diffusivity, coupled with the invariance of the stress peak position with $\varphi$, as before, points to a complex correlation between transient super-diffusion and the stress overshoot. To this end, we also note that while the transient super-diffusion becomes more apparent with increasing $\varphi$, Fig. 8(b) shows that the strength of the stress overshoot deduced from BD simulations and rheometry experiments at low and intermediate Pe (0.1 and 1) exhibits a non-monotonic dependence on volume fraction. The strength of the stress overshoot reduces with volume fraction at high Pe in agreement with previous findings [49]. However at low Pe (0.1 and 1) the strength of the stress overshoot increases up to a critical $\varphi$ around $\varphi$=0.61 for BD and 0.59 in experiments and then decreases well in the glassy regime as random close packing is approached. As Fig. 8(b) shows the increase is more evident for the lower Pe (0.1) suggesting that Brownian motion plays an important role.

In Figs. 17(c) and (d), the non-Gaussian parameter, $a_{2z}$, is shown for the conditions and data sets of 17a and 17b respectively. For the state at rest, $a_{2z}$ shows a general increase with time and with a peak detected in lower volume fraction liquid samples. The latter is detected around the α-relaxation time, therefore within the glass it shifts to much longer times well outside the time window shown here. On the other hand at short times $a_{2z}$ increases with $\varphi$, while at longer times it first increases approaching the glass transition and then decreases at the higher $\varphi$'s well inside the glass, in agreement with previous studies [5, 76].

Under shear $a_{2z}$ shows a peak value and a reduction to zero at long times when the system flows and particle displacements become diffusive. The $\varphi$ dependence is simpler than that at rest with the peak value shifted to lower strains/times, deviating from the state at rest at short times and becoming stronger as $\varphi$ is increased. Hence, at short times the non-Gaussian behavior is more pronounced due to constriction effects (Fig. 17(a)) which hinders in-cage motion, an effect that becomes more important with increasing $\varphi$ and similarly (although not shown) increasing rate.

The $a_2$ parameter is generally associated with dynamical heterogeneities which under shear are detected to be large at intermediate time scales and diminish at long times where shear-induced diffusion sets in. However, this parameter does not fully represent dynamical heterogeneities at all time scales. In the case of large values of $a_2$ at short times, it rather reflects the deviation of particle motion from Gaussian statistics, but simple spatial analysis of dynamic properties [5, 76] shows that it does *not* indicate the existence of dynamic heterogeneities. In the case of short-time constrictions, deviations occur due to the shear-induced cage distortion and refer only to in-cage motions. Thus care should be taken when using the non-Gaussian parameter in relation with dynamic heterogeneity, especially shown for systems under external fields, as previously pointed out in another situation where an external light field was imposed [77, 78].

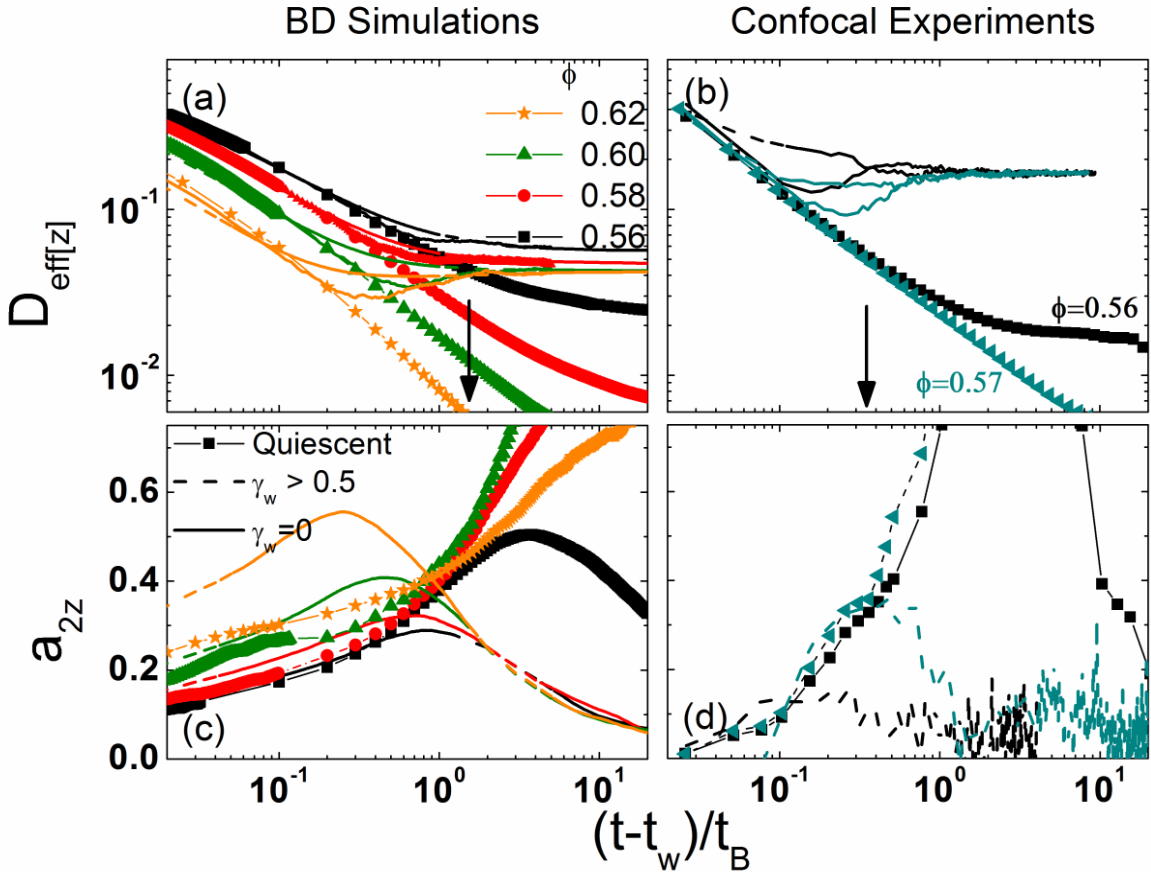

Fig. 17: Two time mean effective diffusivity in the vorticity direction, $D_{eff[z]}=<\Delta z^2>/[2(t-t_w)]$, for various φ in comparison to the state at rest, examined for (a) simulations at Pe=0.1 and (b) experiments $Pe_{sc}$ = 0.594 (φ= 0.56) and $Pe_{sc}$ = 0.652 (φ= 0.57), using waiting strains of $\gamma_w$=0 (transient) and $\gamma_w$>0.5 (steady state). The corresponding non-Gaussian parameter $a_{2z}$ for the same data is shown for (c) simulations and (d) experiments. Arrows in (a) for γ≈0.15 and (b) γ≈0.22, show the approximate stress peak positions for all φ. Time scales for confocal are scaled.

## IV. OVERALL PICTURE AND DISCUSSION

Correlating the transient microstructural information with the stress and particle displacements a comprehensive picture of the transient start-up shear flow emerges. At small strains (≈1%) the stress increases steadily and the structure shows a small distortion, while the displacements following those at rest, an indication of a linear deformation regime. At higher strains (≈10%), within the nonlinear stress response (Fig. 2), particle displacements indicate a transition from in-cage to out-of-cage motion and transient super diffusivity (Fig. 16). g(r) reflects this through stronger intensity of the first maximum in the compression axis (constriction) and a corresponding decrease in the extension axis (Figs. 10 and 11). At even higher strains (≈20%), corresponding to the stress overshoot, the long-time displacements are already almost diffusive. The xy plane g(r) is consistent with this behavior, with the compression axis of the first maximum showing higher values, while indicating diffuse particle locations along the extension axis due to a large contribution from out-of-cage motion (Figs. 10 and 11). At the steady state regime (>60%), the measured stress relaxes from its peak value (≈20%), while particle displacements are diffusive over long distances and structures have reached a final steady state which still is distorted.

While the stronger constriction detected in the compression axis corresponds to a higher stress with increasing strain due to cage elasticity, the out-of-cage motion should reflect viscous losses. Thus in the linear regime and up to 10% where there is little or no out of cage diffusion, the elastic behavior dominates with a strong increase of stress with strain. As out-of-cage rearrangements become more dominant, the stress peak is reached with cages almost reaching maximum distortion. When shear is comparable to the Brownian motion (Pe=1), some particles remain trapped within cages even though maximum distortion has been attained. Consequently a stress drop occurs as the strain increases further, associated with the release of particles that were still trapped within the cages at the peak stress, allowing a relaxation of stored stress.

Therefore the mechanism behind the stress peaks lies in the persistence of structural cage elasticity, since the maximum cage distortion occurs before all particles are able to undergo out of cage relaxation. This intuitively denotes that high shear leads to more elongated cages during the stress overshoot, which are partly or even totally destroyed at the steady state. This argument is strengthened by the experimental and simulations finding that smaller rates and higher φ's produce smaller stress peaks. The former is a matter of particles having more time to diffuse out of their cage with Brownian motion thus allowing the system to relax with less structural distortion, while the latter occurs because as φ is increased, less available space makes any cage distortion more difficult. This picture holds mainly at lower volume fractions and Pe whereas well in the glass regime the overshoot initially shows a non-monotonic behavior with Pe and as φ approaches rcp shows a decay with increasing Pe. In this regime shear induced collisions dominate over Brownian motion (the system is approaching the non-Brownian regime) and entropic elasticity is no longer an efficient mechanism for cage relaxation.

Through the use of MCT, the work of [46] connects the appearance of the stress peak to the local transient super diffusion and a negative value of a generalized dynamical shear modulus G(t). In their formalization of G(t), they connect the decrease of stress after the peak to an approximate time dependent isotropic structure factor. Although we found the connection of the stress peak to the transient super diffusion to be quite complex, our structural data during and after the peak do show strong correlation to the transient stress response. Unlike the theoretical approximation, the structural information extracted from the simulations is highly anisotropic (Fig. 12). Some of the details of the stress-strain dependence and the related underlying microstructural and dynamic behavior seen in experiments and BD simulation are not captured by MCT models. For example MCT clearly overestimates the values of the strain where the stress overshoot takes place, although the increase with Pe is qualitatively captured [44], and MCT can not account for the diminishing stress overshoot approaching random close packing or with increasing Pe in the high Pe regime. Similarly the difference in the position of the stress overshoot and the onset of super-diffusive behavior is not considered by the theory. Therefore among other possible origins for such discrepancies that would be interesting to explore is the use of an anisotropic structure as an input to MCT models in order to compare the predictions with data from experiments and simulations.

The stress, structure and microscopic dynamics under shear presented here, might provide insights on the connection between thermal and shear activated cage melting and therefore the comparison between shear rejuvenation and thermally activated equilibration as suggested in literature [79, 80]. However, from another perspective, the introduction of thermal and shear energy to a system may not be equivalent. Whereas as it was shown by [47], shear-induced diffusivity measured by confocal microscopy under shear may not increase linearly with shear rate as it does with temperature, recent experiments show that the relaxation time measured by orthogonal superposition rheometry follows a linear dependence on shear rate [75].

Note that shear induced diffusivities here depend on the direction (although not significantly), with higher mobility occurring in the extension axis than the compression axis.

However, long-time shear-induced and thermal motions are found to be diffusive showing Gaussian displacements [46, 47]. Furthermore one might expect that even if anisotropy in shear-induced particle displacements is present, thermal and shear energy input may lead to similar effects after shear cessation and may therefore be equivalent in terms of erasing material memory [79]. Nevertheless, experiments and simulations suggest that shear induces structural anisotropy which might not relax (at least within reasonable experimental time) to isotropic ones in metastable glassy states and therefore are causing residual stress remaining in the system for a long time after the shear is switched off [81].

## V. CONCLUSIONS

We have presented transient stresses, structures and particle displacements during a start-up shear experiment in concentrated hard sphere colloids with experimental rheology, Brownian Dynamics simulations and confocal microscopy. We find that the strength of the stress overshoot in experiments and simulations increases with increasing Pe up to a characteristic Pe that is decreasing with $\varphi$. The yield strain related with the stress peak increases with Pe and remains generally unchanged with $\varphi$, in both experiments and simulations. The relative height of the stress peak as well as the strain is associated with the structural elongation of the cage during shear, as evidenced by the anisotropy of the pair correlation function at contact. The transient g(r) in the xy plane at low strains is isotropic, turning into a distorted structure as strain increases and showing flow lines along the extension axis when strains increase above yield signifying out of cage diffusion. While cage distortion is more significant at high rates, low rates still distort the structure, although do not elongate the cage enough to give rise to a stress peak.

Analysis of particle displacements additionally shows that yielding is initiated along the extension axis, verifying the structural results. Moreover we find that long-time diffusivities at steady state are slightly different in various directions, with the highest in the extension axis, followed by the compression and last the vorticity axis (as deduced from Figs. 16 and 17). Simulations also show that short-time displacements under shear decrease, an effect which is more pronounced with increasing $\varphi$ and Pe. These constrictions can be associated with the shear induced out-of-cage motion, as the hard-sphere hindrance due to a progressively deformed cage actively forces particles to move from the compressional to the extensional direction leading to particle rearrangements and cage breaking with a shear-induced long-time diffusivity that increases with shear rate.

A transient super-diffusive regime occurs at intermediate time scales, probing particles escaping their cages during start-up flow, an effect that is more pronounced with increasing $\varphi$ as cages become tighter. Although the stress overshoot and the super diffusivity occur through the yielding process, they are not qualitatively correlated. High Pe interestingly shows a steady state super diffusive regime, which relates to the short-time shear-induced cage constriction. Steady state super-diffusion in conjunction with simple long-time diffusivity point to a state under shear where particle cages are constantly broken and reformed.

## ACKNOWLEDGMENTS

We acknowledge funding by Greek project Thales "Covisco" and Aristeia II "MicroSoft". N.K. has been supported by EU Horizon 2020 funding, through H2020-MSCA-IF-2014, ActiDoC No. 654688. We thank the Deutsche Forschungsgemeinschaft (DFG) for support through the Research Unit FOR1394 (Project P2).